\documentclass[conference]{IEEEtran}
\ifCLASSOPTIONcompsoc
\usepackage[nocompress]{cite}
\else
\usepackage{cite}
\fi
\ifCLASSINFOpdf
 \usepackage[pdftex]{graphicx}
 \usepackage{dblfloatfix}
\else
\fi
\usepackage{amsmath}
\ifCLASSOPTIONcompsoc
  \usepackage[caption=false,font=footnotesize,labelfont=sf,textfont=sf]{subfig}
\else
  \usepackage[caption=false,font=footnotesize]{subfig}
\fi
\usepackage{url}

\usepackage{multirow}
\usepackage{flushend}
\begin{document}
\title{Practical Decentralized Attribute-Based Delegation using Secure Name Systems}

\author{\IEEEauthorblockN{Blinded for review}}
\author{\IEEEauthorblockN{Martin Schanzenbach, Christian Banse and Julian Sch\"utte}
	\IEEEauthorblockA{Fraunhofer AISEC, Germany\\
		Garching near Munich, Germany\\
		\{schanzen,banse,schuette\}@aisec.fraunhofer.de}}
\maketitle

\begin{abstract}

Identity and trust in the modern Internet are centralized around an oligopoly of identity service providers consisting solely of major tech companies. The problem with centralizing trust has become evident in recent discoveries of mass surveillance and censorship programs as well as information leakage through hacking incidents.
 One approach to decentralizing trust is distributed, attribute-based access control via attribute-based delegation (ABD).
Attribute-based delegation allows a large number of cross-domain attribute issuers to be used in making authorization decisions.
Attributes are not only issued to identities, but can also be delegated to other attributes issued by different entities in the system. 
The resulting trust chains can then be resolved by any entity given an appropriate attribute storage and resolution system. 
While current proposals often fail at the practicability, we show how attribute-based delegation can be realized on top of the secure GNU Name System (GNS) to solve an authorization problem in a real-world scenario. 
\end{abstract}


\section{Introduction}\label{sec:Introduction}

Communication paradigms are shifting from centralized client-server architectures to decentralized communication between interconnected devices and services. 
Trends like the Internet of Things (IoT) or technologies like Blockchain are only some examples in which peers directly interact with each other across trust boundaries, as opposed to traditional content mediation by central trusted services. 
The decentralization of communication and the diversification of trust domains comes with major challenges on identity and access management (IAM). 

In traditional centralized architectures within a single trust domain, IAM is well understood and various solutions are available. 
They typically consist of a trusted party that manages authentic descriptions of entities in the system. 
This information serves as the trusted foundation for access control models all built on the assumption that attribute information is managed by a central trusted instance. 
In decentralized architectures, this assumption does not hold any more, as there is no central authority with ultimate trust that would be able to manage attributes and guarantee their authenticity.


The architectures we see in today's applications often try to build around this problem rather than solve it. 
One workaround is to introduce a central trusted third party that is responsible for managing attribute information, although it might not be needed from an architectural point of view. 
The negative consequences are impressively demonstrated by the growing number of platform ``silos'' in the IoT where originally decentralized architectures are turned into centralized closed worlds of platform vendors. 
Besides the fact that such centralization is not economically desirable, as it prevents the creation of truly interoperable and open systems, it has a massive impact on users' privacy. 
A central identity provider keeps all attributes of all entities in the system, is aware of any user activity and is thereby able to apply behavioral analysis.
Most web-based applications today rely on only two identity providers, namely Google and Facebook~\cite{website:gogyaidmarket}, which creates a de facto oligopoly of omniscient central parties. 
Another workaround is identity federation, i.e. the attempt to make IAM systems of multiple trust domains interoperable with each other. 
Examples are cross-certification of certificate authorities in a PKI or protocols like SAML for exchanging identity information between domains. 
While this approach generally works, it comes at significant overhead costs and still does not address privacy problems of central identity providers.

Alternative approaches for a decentralized management of attributes have been proposed in literature before, using attribute-based delegation (ABD) ~\cite{blaze1998compliance,blaze1999keynote,clarke2001certificate} -- a technique to delegate individual attributes from one principal to another in a decentralized way. 
ABD has been shown in prior research \cite{blaze1996decentralized,lee2008towards} to be a suitable solution to model complex, decentralized trust relationships. 
Access control based on delegated attributes eliminates the need for a central IdP and allows for scalable, decentralized administration of attributes and, consequently, authorization. 
Standardized authorization frameworks such as UMA could benefit from ABD-based policies, allowing trust relationships to exceed the boundaries of the UMA server. 
However, research on ABD has so far been mostly theoretic and left questions, such as the practical applicability, out of scope.

In this paper, we build upon the existing concept of ABD and present a practically usable, decentralized authorization system. 
Specifically, we extend existing ABD schemes and show that their implementation can be achieved efficiently on the basis of existing secure name systems, which, by their nature, provide the capabilities and security properties required for ABD schemes. 

In summary, this paper makes the following contributions:

\begin{itemize}
	\item The finding and argumentation that secure name systems can be used as a basis for ABD schemes
  \item Design of a practical, decentralized ABD system based on a secure name system
	\item Implementation of a prototype on top of the GNU Name System (GNS) for a real problem scenario
\end{itemize}

To the best of our knowledge, we are proposing the first practically usable system for decentralized, attribute-based delegation. 

\section{Related Work}
\label{sec:RelWork}
In this section, we discuss existing approaches to attribute-based delegation including some of their shortcomings.
According to Lee et al~\cite{lee2008towards}, authorization in decentralized environments can be categorized in \emph{distributed proof} and \emph{trust negotiation} techniques. 
ABD systems are distributed proof techniques.

One example for an ABD system is the Simple Distributed Security Infrastructure (SDSI/SPKI)~\cite{rivest1996sdsi} by  Rivest et al. 
In SDSI, authorization is obtained when a collection of user attributes is in compliance with a requested resource security policy. 
SDSI implements ABD by defining credentials as authorization certificates called \emph{auth certs}. 
In SDSI, $A.a \rightarrow B$ denotes that an issuer with public key $A$ grants the attribute $a$ to the subject with public key $B$. 
$B$ may itself act as an issuer and further delegate the attribute $a$. 
In addition to the attribute delegation, SDSI allows for the controlling of the depth of delegation, limiting the number of times an attribute may be delegated. 
For chain discovery, Clarke et al~\cite{clarke2001certificate} have proposed an algorithm that treats a delegation as a ``rewriting'' rule and discovery as a term-rewriting problem. 
Term-rewriting is performed until no more rewriting rules can be applied or an attribute that satisfies a policy is found.
However, the algorithm relies on the availability of a complete set of attributes prior to chain discovery. 
In the case of changes in the context of the system and revocation or expiration of the credentials, the chain-discovery algorithm may not be deterministic. 
This is a requirement that is rather difficult to ensure in practice, especially in distributed scenarios. 
It is much more likely that small subsets attributes and credentials need to be resolved and verified on demand to make authorization decisions. 
Resolution and distributed storage of attributes and credentials is considered out of scope in SDSI.

Li et al~\cite{li2003rt} proposed an authorization mechanism based on ABD as well as a trust management language called $RT_{0}$. 
It allows one to express different types of credentials as attributes. 
All credentials can be represented in a \emph{credential graph}. 
Finding a path in a credential graph is equivalent to finding a credential chain as a proof of authorization. 
$A.a \leftarrow B.b$ is an example of a credential in $RT_{0}$. 
It also supports SDSI delegations, since the auth certs in SDSI can be translated into $RT_{0}$ with the exception that the arrows are reversed in the notations. 
There are four different types of attribute delegations:

\setcounter{equation}{0}
\begin{align}
  A.a \leftarrow & B\\
  A.a \leftarrow & B.b\\
  A.a \leftarrow & B.b.a\\
  A.a \leftarrow & \bigcap_{i=1}^n f_i
\end{align}
Delegation type 1 is interpreted as ``$A$ calls $B$ $a$''. 
The type 2 delegation $A.a \leftarrow B.b$ is interpreted as ``All entities that $B$ calls $b$ are called $a$ by $A$''. 
The type 3 delegation $A.a \leftarrow B.b.a$ is interpreted as ``All entities that are called $a$ by all entities that $B$ calls $b$, are also called $a$ by $A$''. 
The type 4 delegation $A.a \leftarrow \bigcap_{i=1}^n f_i$ is interpreted as ``$A$ calls all entities $a$ that satisfy each $f_i$'' with $f_i$ being any right-hand expression of either (1),(2) or (3). 

The chain-discovery algorithm to build and traverse a credential graph proposed by Li et al~\cite{li2003distributed} for $RT_0$ unifies a backward search from issued attributes to the subjects and a forward search from subjects to the issued attributes to a bidirectional search. 
The authors argue that this approach finds credential chains even if they allow credentials to be stored with either issuers or subjects in distributed scenarios. 
However, the algorithm guarantees chain discovery \emph{only} by enforcing strict and complex constraints on the credential storage to ensure what they call ``well-typedness''. 
Additionally, the worst case time complexity of the algorithm for $N$ credentials is $O(N^3)$.

The above approaches both propose partial solutions to solve the problem of making authorization decisions in distributed or decentralized scenarios. 
In this work, we show that the existing research in the area of ABD can be used in combination with secure name systems to create a practically usable ABD by addressing the issue of delegation storage and resolution in a decentralized enviroment.

\section{Scenario}
\label{sec:abd}
To motivate and demonstrate our approach we present the following real-world scenario: 
A startup company wants to improve the anti-doping process by providing a secure and privacy-preserving service $S$ to athletes as well as anti-doping organizations. 
$S$ allows official doping control officers (DCOs) to request the current location of an athlete in the field. 
We will not address the privacy implications that such a service must also consider, but focus on the organizational authorization needs in this particular scenario. 
National anti-doping organizations (NADOs) are organizations that incorporate and adhere to the world anti-doping code by the World Anti-Doping Agency (WADA). 

While NADOs can be recognized by WADA for adhering to the code, there is no hierarchical relationship between them. 
NADOs are responsible for organizing and executing doping tests for athletes in their respective regional domain, but then often delegate the actual controlling to subcontractors.
Note that although the organizational structure might suggest a hierarchical relationship between the entities, this does not imply a central management of authorization attributes. 
WADA is neither interested in nor authorized to manage attributes of DCOs. It may simply assert that a NADO does adhere to the code. 
One might be tempted to resort to traditional PKIs and use NADO sub-CAs for delegating attribute management from a central WADA CA to NADOs. 
However, in particular X.509 is limited by design to bind a key to a subject in which the subject is uniquely identified by a globally unique name. 
Additionally, it does not directly address attribute delegation and resolution, making issuing and revoking attributes at runtime tedious processes. 
Rather, this scenario highlights the need for decentralized attribute management: The service $S$ wants to authorize subjects based on attributes that it delegates to all entities considered to be DCOs by NADOs.
This can easily be modeled using ABD in a formal way.
We use Li's notation including the operator $\leftarrow$ to denote attribute delegations. 
On the left side of the operator, we write the \emph{issuer} and the \emph{delegated attribute} and, on the right side, the \emph{delegation subject expression}.
Note that according to Li et al.~\cite{li2003distributed} in a type 3 delegation $A.a \leftarrow B.b.a$ the second $a$ of $B.b.a$ must be the same attribute that is specified on the left side of the expression. 
Additionally, it is not allowed to specify arbitrarily long delegations on the right-hand side of the expression. 
However, this limitation is unnecessary and solely introduced by them to simplify the proposed algorithm. 
We can lift this restriction by imposing \emph{issuer-side storage} of delegations. 
Issuer-side storage refers to attributes that are managed and stored by the issuer itself. 
The opposite approach is called \emph{subject-side storage}, in which the subject manages and stores the issued attribute. 
Consequently, delegations in the form $A.a \leftarrow B.b_1.[...].b_n$ are perfectly acceptable in our design.
We model delegations in the aforementioned scenario as follows:

\setcounter{equation}{0}
\begin{minipage}[t]{0.9\columnwidth}
  \begin{small}
  \begin{align}
    \mathit{S}.user &\leftarrow \mathit{WADA}.nado.dco\\
    \mathit{WADA}.nado &\leftarrow \mathit{NADA}\\
    \mathit{WADA}.nado &\leftarrow \mathit{USADA}\\
    \mathit{NADA}.dco &\leftarrow C_1.dco\\
    \mathit{USADA}.dco &\leftarrow USADA.contractor.dco\\
    \mathit{USADA}.contractor &\leftarrow C_2\\
    \mathit{C_2}.dco &\leftarrow{C_2.employee}~\cap \nonumber\\
    &\quad\enspace{C_2.controller}
  \end{align}
\end{small}
\end{minipage}
\begin{minipage}[t]{0.964\columnwidth}
  \begin{small}
  \begin{align}
  C_1.dco &\leftarrow Alice\\
  C_2.employee &\leftarrow Bob\\
  C_2.controller &\leftarrow Bob
  \end{align}
\end{small}
\end{minipage}


In (1), the service $S$ delegates the attribute $user$ to all entities that have the attribute $\mathit{WADA}.nado.dco$. 
$\mathit{WADA}$ itself delegates the attribute $nado$ to all national anti-doping organizations that adhere to the world anti-doping code, such as the German ``Nationale Anti Doping Agentur'' $\mathit{NADA}$ (2) and the ``U.S. Anti-Doping Agency'' $\mathit{USADA}$ (3). 
NADOs then delegate the attribute $dco$ to their subcontractors. 
$NADA$ subdelegates the $dco$ attribute to the contractor $C_1$ (4), while $USADA$ uses a dynamic attribute $contractor$ (5) to define all subcontractors that presently have a control assignment (6). 
This attribute is revoked or will expire as soon as the assignment ends. 
The subcontractors may either delegate the $dco$ attribute to an attribute expression that is more meaningful to the contractor (7) or directly assign it to an entity (8).

We can make the observation here that $\mathit{WADA}$ is not necessarily aware of the delegation in (1) and most likely not even interested in this information. 
In traditional, centralized scenarios, the service $S$ would allow $\mathit{WADA}$ to issue the attribute $user$ on its behalf. 
However, in decentralized scenarios, the lack of a trust or organizational relationship between two parties is more common than not. 
In this case, the delegations are always stored with the issuer and never with the subject.
The only time it is reasonable to store a delegation with a subject, is when this information is useful \textit{and} meaningful to the subject itself. 
For example, the delegations in (8,9,10) are stored with $Alice$ and $Bob$, respectively, for two reasons: First, both $Alice$ and $Bob$ know their employers and are aware that they are employees. 
Second, $Alice$ and $Bob$ use this information regularly to prove that they are actually employees of $C_1$ and $C_2$, respectively. 
Based on this observation, we define two types of attributes in an ABD system: issuer-stored \textit{Delegations} (1-7) and subject-stored \textit{Credentials} (8,9,10).

\section{Design}
In this section, we present our design of a decentralized, attribute-based delegation system on top of a secure name system. 
Secure name systems, such as the Domain Name System (DNS) with Security Extensions (DNSSEC)~\cite{RFC4033}, namecoin\footnote{\url{https://namecoin.info/}, accessed 5/19/2017} or the GNU Name System (GNS)~\cite{wachs2014feasibility,wachs2014censorship}, provide a secure mapping from attributes to resources. 
Resources can be publicly queried by all peers, but creation and updates are only possible by their respective owners. 
Name systems are agnostic towards the interpretation of attributes and values and rather serve as a distributed management and discovery mechanism.

Name systems consist of \emph{namespaces} that are owned by private or legal entities. 
Namespaces are managed by their respective owners and contain name-value mappings. 
We consider name systems as a suitable basis for the implementation of an ABD system because of the possibility for an owner to delegate the authority over names to namespaces of other owners. 
In the context of ABD, the owner of a namespace is an ``issuer'' of attributes or attribute delegations in its respective namespace. 

The owner of a namespace specified as a value in a delegation is a ``subject''.
As such, name systems inherently provide a storage, resolution and delegation mechanism for issued attributes and their delegations.
Name-value mappings are realized in name systems using \emph{resource records}. 
The content of a resource record is defined by a \emph{type}, such as ``A'' for the most common record in the Domain Name System (DNS): An IPv4 address.

In the following, we present our ABD design using \emph{Attribute Delegation Records}, how delegation chains can be resolved using \emph{Delegation Chain Discovery} and an authorization flow using ABD. 
Finally, we briefly discuss attribute revocation as well as security and privacy implications of different secure name systems.



\subsection{Attribute Delegation Records}

We introduce a special resource record type ``ATTR'' for attribute delegations and modify the name system resolver logic to perform delegation chain discovery for such records.
Attribute delegations such as $A.a \leftarrow e$ as introduced in Section~\ref{sec:abd} are mapped into a namespace as follows:
$A$ is a namespace owned by an entity and $a$ is the name of a record in $A$. The value of the record contains $e$, the delegation expression that defines the namespaces that $a$ is delegated to.
To support all four kinds of attribute delegations, our record contains an appropriate data structure to hold any attribute expression $e$ in a delegation $A.a \leftarrow e$.

Specifically, we define the value of an ``ATTR'' record to contain one or more entries in a \textit{delegation set}. 
A delegation set entry consists of a subject namespace $B$ as well as a set of attributes and is used to represent delegation types 1-3.
To model delegation types 1-3, a resource record contains a single entry in the delegation set. 
A type 4 delegation record contains a delegation set with $n$ entries, each specifying the respective required delegation expression $A.a \leftarrow \bigcap_{i=1}^n f_i$.

\begin{figure*}[t]
  \centering
  \subfloat[Namespaces.]{\label{fig:scenario_namespaces}\includegraphics[width=0.9\columnwidth]{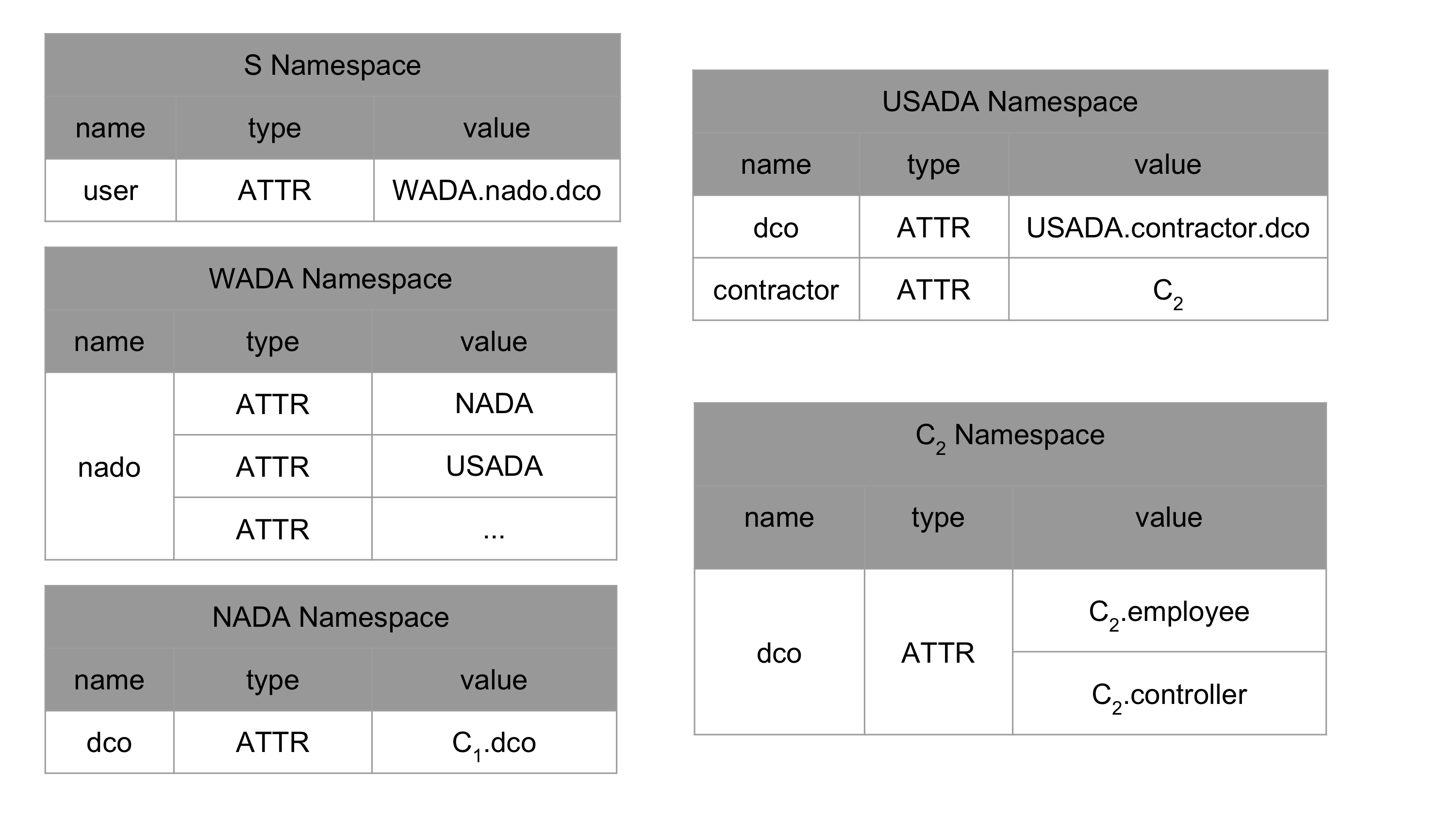}}~
  \subfloat[Delegation graph.]{\label{fig:delegationgraph}\includegraphics[width=1.1\columnwidth]{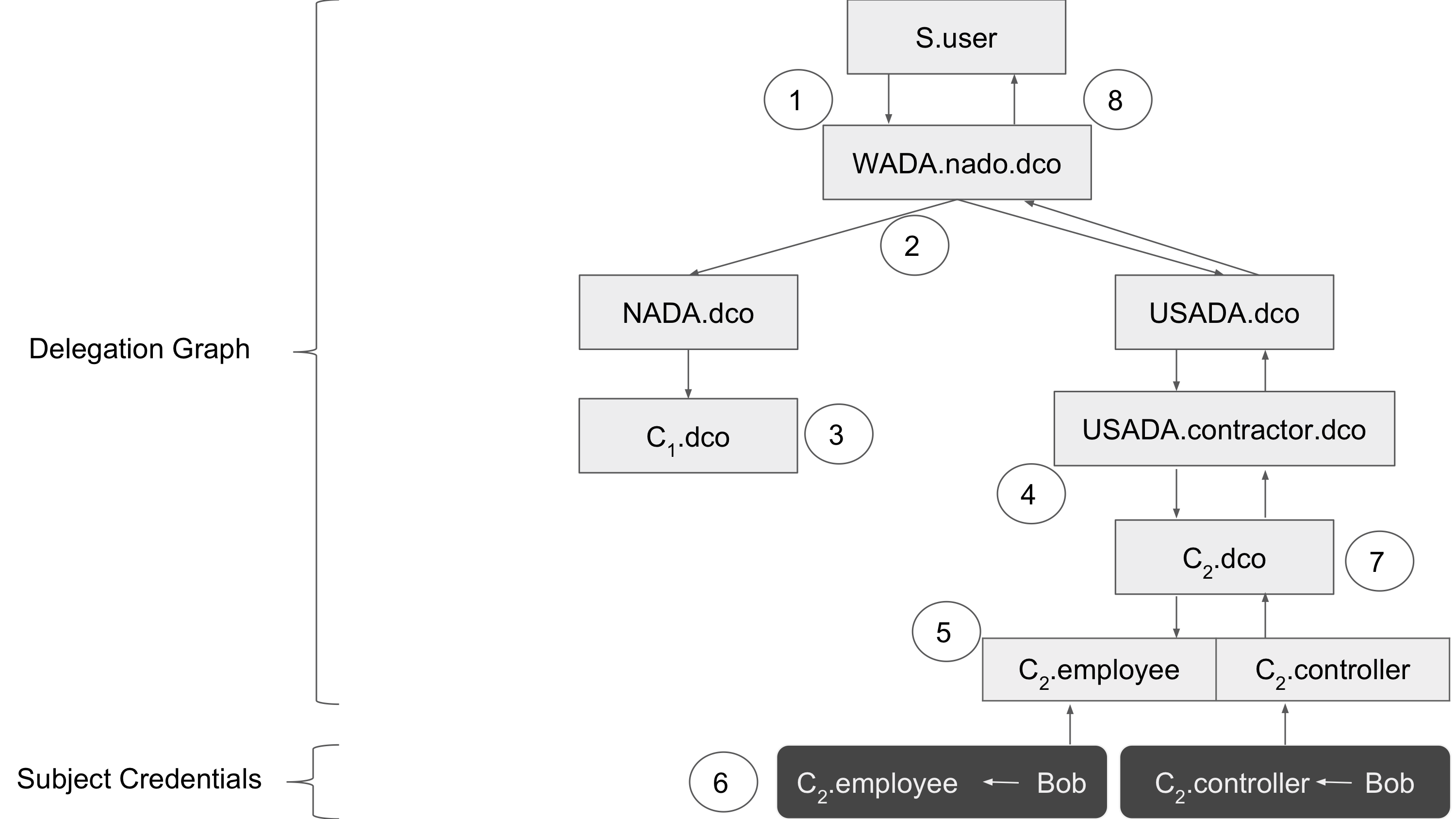}}
  \caption{Delegations in the reference scenario from $S.user$ to $Bob$.}
  \label{fig:ref_scenario}
\end{figure*}

While the type 4 delegation constitutes a logical ``AND'', a logical ``OR'' is not explicitly defined. 
However, the existence of multiple delegation records in the same namespace under the same attribute $a$ implicitly defines this case. 
In Figure~\ref{fig:scenario_namespaces}, the namespaces for our reference scenario are illustrated. The namespace of contractor $C_1$ does not have any delegations so it is omitted. 
Representing a type 1 delegation, the $\mathit{WADA}$ namespace contains multiple records under the name $nado$ with a single delegation set entry. 
The type 4 delegation in the $C_2$ namespace contains only a single record under the name $dco$ with two delegation set entries.


\subsection{Credentials}
We assume subjects are issued attribute-based credentials (ABCs) by a variety of issuers including employers, email providers, or nation states. 
In our design, a subject $B$ has a set of ABCs $C_B$. 
Each credential $c \in C_B$ may be issued by a different issuer and asserts the user an attribute.
In theory, credentials could be represented as type 1 attribute delegations.

However, this is disadvantageous for two reasons: First, by having all credentials in the delegation system it is possible for anybody to enumerate all entities that have credentials for a specific attribute.
Second, we want our system to support all kinds of credentials, including privacy-preserving, attribute-based credentials (PP-ABCs), such as \cite{camenisch2002design} or \cite{paquin2013u}, that cannot be persistently stored and have to be presented online.
Therefore, our design is agnostic to the representation of the credential.

\subsection{Delegation Chain Discovery}
\label{sec:chain_discovery}
To confirm that an issuer delegated an attribute $a$ to an entity $B$, a delegation chain must be discovered. 
A valid chain can be found if $B$ holds a set of credentials $C_{A.a} \in C_{B}$ that allows one to build a delegation chain $D_{A.a,B}$ from an issuer namespace $A$ and attribute $a$ to $C_{A.a}$.

Finding a delegation chain can only be guaranteed if all attribute delegations $d \in D_{A.a,B}$ are resolvable \textit{and} $B$ is in possession of an appropriate set of credentials $C_{A.a}$.
We define the resolver function $resolve(l,N,t)$ that is used to resolve resource records of type $t$ under the name $l$ in the namespace $N$.
A call to $resolve(a,A,``ATTR")$ will return the resource records representing all issued attribute delegations $A.a \leftarrow e$ as delegation sets. 
Each expression $e$ in the delegation sets is checked against the issued attributes in the set of subject credentials $C_{B}$. 
If we have found a valid delegation chain from the original attribute to a credential subset $C_{A.a}$, we have verified that the attribute $A.a$ is delegated to $B$. 
Our algorithm is a combination between SDSI-style rewriting and the backward resolution of a delegation graph in Li's approach for $RT_0$. 
However, as we enforce issuer-side storage by defining delegations in the issuer namespace, we do not require the more complex unified approach by Li that uses backward \emph{and} forward search of the delegation graph.

To resolve a delegation $D_{A.a,B}$ using a name system, the namespace of the issuer and the attribute to look up must be known in advance.
For an initial attribute $A.a$ the name to look up is $a$ in the namespace $A$. 
We define $A.a$ as the root node and all resolved delegation expressions $e$ found under $A.a$ as children of $A.a$ in the delegation graph. 
From then on, we follow a rewrite-resolve-check pattern until we can match a credential against a delegation subject.
If a resource record containing a delegation set with a single entry is resolved, the expression $e$ is of type~1-3. Otherwise, it is of type~4. In both cases, we use SDSI-style rewriting rules~\cite{clarke2001certificate}:

For a type~1-3 delegation $e := B.b_1.b_2.b_3...b_n$ we perform a lookup query using only the leftmost attribute $b_1$ and we \textit{rewrite} the resulting expressions from a call to $resolve(b_1,B,``ATTR")$ by appending $b_2$ through $b_n$. 
This leads to a reduction of the original delegation expression if the query returns a type~1 delegation or an enlargement for a type~3 delegation. 
In the case of a type~2 delegation, the expression complexity does not change.
For a type~4 delegation $e := \bigcap_{i=1}^{n} f_i$ we process each $f_i$ like type~1-3. Rewriting a type~4 delegation set is simply a matter of rewriting every delegation set entry individually.

The rewritten delegations are added as children of $e$ in the delegation graph and checked against the subject credentials. 
The process continues iteratively until a matching set of credentials is found that allows us to backtrack the delegation graph to $A.a$. 
When we backtrack the delegation graph and encounter a node that holds a type 1-3 delegation, it is verified that the delegated attribute has a path to a set of subject credentials and we can continue backtracking. 
If we encounter a node representing a type 4 delegation, we have to make sure that every $f_i$ in the node is satisfied by a set of credentials before we can continue. 

Figure \ref{fig:delegationgraph} illustrates a delegation chain discovery for our scenario described in Section \ref{sec:abd} and the namespaces in Figure~\ref{fig:scenario_namespaces}: (1) $S.user$ is resolved to a single record with the delegation set entry $\mathit{WADA}.nado.dco$, a type 3 delegation.
(2) $\mathit{WADA}.nado$ resolves to two records containing one delegation set entry each: $\mathit{NADA}$ and $\mathit{USADA}$. The rewritten expressions $\mathit{NADA}.dco$ and $\mathit{USADA}.dco$ are added to the graph.
(3) $\mathit{NADA}.dco$ resolves to a record containing the delegation set entry $C_1.dco$. Bob does not have a credential to satisfy this delegation.
(4) $\mathit{USADA}.dco$ resolves to $USADA.contractor.dco$. The dynamic attribute $USADA.contractor$ resolves to $C_2$ leading to $C_2.dco$.
(5) $C_2.dco$ resolves to a single record containing two delegation set entries representing the type 4 expression $C_2.employee~\cap~C_2.controller$.
(6) Bob has the credential $C_2.employee$ that matches the delegation set entry $C_2.employee$. The graph is backtracked but the delegation set containing $C_2.controller$ is not yet fulfilled. Bob's credentials are checked against $C_2.controller$ and the credential $C_2.controller \leftarrow Bob$ satisfies the delegation set in (5).
(7) The delegation graph is backtracked further until $S.user$ is reached and the delegation chain is successfully discovered in (8).

\subsection{Authorization using Attribute-Based Delegation}
\label{sec:abd_auth}
\begin{figure*}[h]
  \centering
  \includegraphics[width=0.9\linewidth, trim={0 4cm 0 0}]{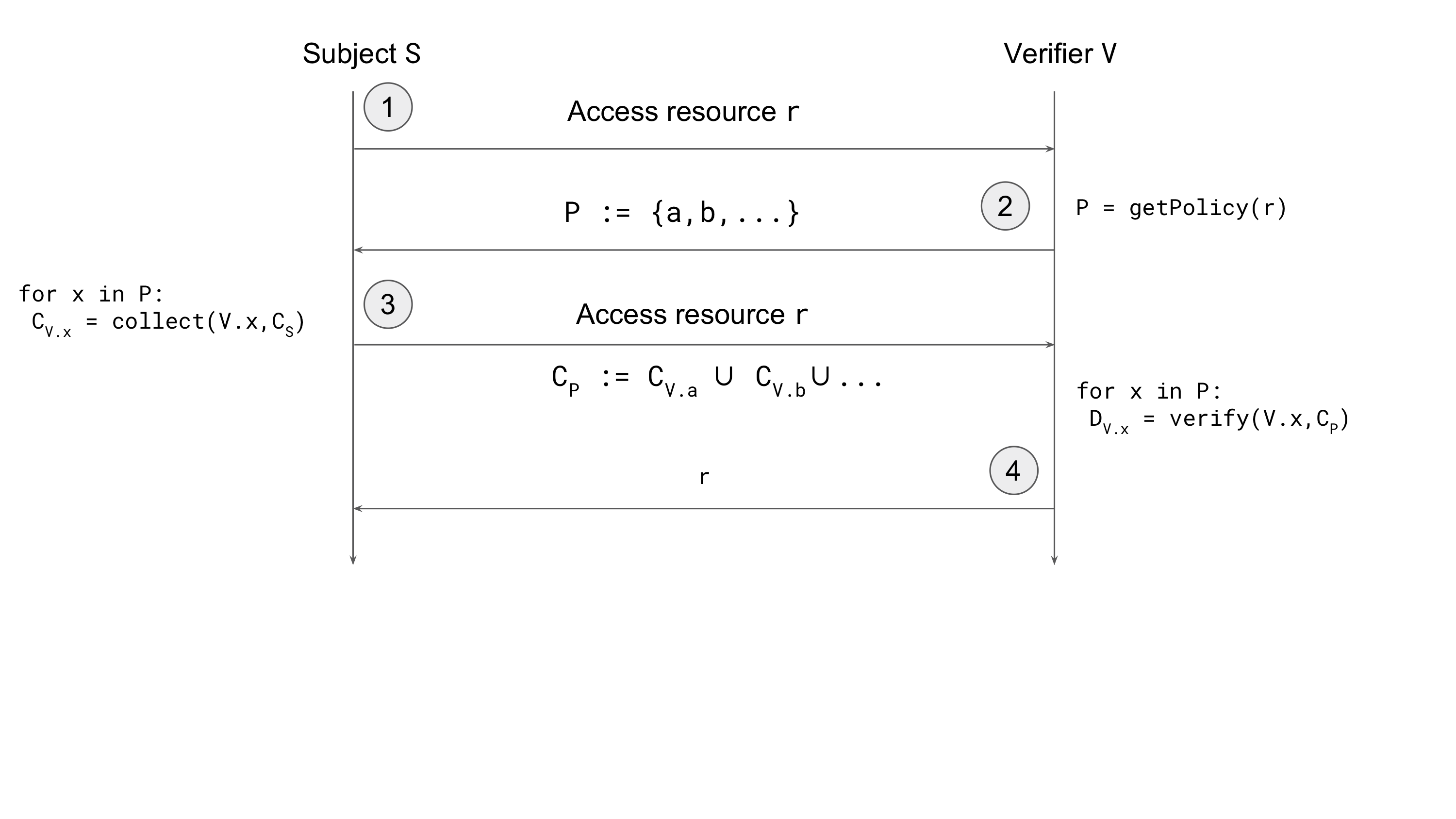}
  \caption{Authorization with delegation chain discovery.}
  \label{fig:negotiation}
\end{figure*}

While we have established how attribute delegations can be resolved and verified above, we now introduce a protocol to actually authorize subjects to access resources protected by policies containing delegated attributes. 
We define a resource $r$ to be protected by a policy $P$ that specifies a set of attributes. 
A verifier $V$ can perform attribute-based authorization of a subject that requests access to $r$. 
To do so, the verifier initially retrieves $P$ by querying a policy storage. 
We define the attribute issuer for all attributes $x \in P$ to be the verifier $V$.
The verifier is initially unaware as to which credentials the subject must provide to satisfy $P$. 
At the same time, the subject is initially not aware of what attributes are required by $P$ to access $r$.
Our simple authorization protocol with delegation chain discovery is illustrated in Figure~\ref{fig:negotiation}.

\begin{itemize}
  \item[(1)] The subject $S$ tries to access the resource $r$.
  \item[(2)]  To retrieve the access policy $P$ for the resource $r$, $V$ uses a function $getPolicy$. Afterwards, $V$ sends a response containing the policy $P$.
  \item[(3)] $S$ uses a function $collect$ that finds subsets $C_{V.x}$ of the subjects credentials $C_S$ that satisfy the attributes $x \in P$. $S$ sends the set $C_P := \bigcup_{x \in P} C_{V.x}$ to the verifier.
  \item[(4)] We define a function $verify$ that uses delegation chain discovery to verify that a delegation chain $D$ exists for a set of credentials to an attribute. $V$ uses this function to confirm that a delegation chain $D_{V.x}$ can be found for all $x \in P$ using $C_P$. Access is granted only if all delegation chains can be found.
\end{itemize}

\subsection{Revocation}
Revocation of a delegation is achieved by having the respective issuer revoke the attribute name that points to it in the name system.
A name in the name system can only be resolved if it exists and is not expired. Attribute delegations must be treated in the same fashion. 
Whenever an issuer wants to remove an attribute delegation, he must delete the respective records from his namespace. 
It is also important that attribute delegations must have a set expiration date. 
Distributed name systems tend to cache records in the network until they expire. 
Even if a record is deleted by the namespace owner, it might still linger until caches are purged or the record has expired. 
Records may have relative expiration times that can be set to a short duration. Such records will be automatically renewed after they have expired in the owner's namespace. 

Revocation of \textit{credentials} is not directly related to the name system. Depending on the used attribute-based credential system, revocation is performed by the issuer and revocation checking must be performed by a verifier to ensure that a provided set of credentials is still valid.
\subsection{Security and Privacy}
While, theoretically, all name systems are suitable for attribute-based delegation as discussed above, practically only name systems with strong security guarantees are reasonable choices when actually building such systems in practice.
An attribute delegation $A.a \leftarrow e$ must be verifiably issued by the owner of $A$ and be resolvable as such. 
An insufficiently resilient name system might be subject to denial of service attacks rendering the ABD systems useless. 
Also, bulk collection and enumeration of attribute delegations is unwanted, as it exposes organizational and/or trust relationships.
For this reason, we will take a brief look at secure name systems and their properties.
We limit our discussion to three designs that represent three different approaches to secure name systems: namecoin, a blockchain-based name system, the GNU Name System (GNS) and DNSSEC, the security extensions for the Domain Name System.
We look at namecoin and GNS because they offer protection against attacks such as client observation on the network and operator level as well as censorship and/or legal attacks~\cite{DPRIV17}. 
DNSSEC, on the other hand, is the most widely adopted secure name system. 
In general, properties of secure name systems include:

\paragraph{Integrity} The integrity of records in a namespace can be ensured by having the namespace owner provide a digital signature along with the resource records. All three name systems follow the same approach.

\paragraph{Availability} Record availability is addressed in namecoin by having all records replicated by all participants. However, Blockchain-based decentralized protocols are still prone to various attacks~\cite{hijackbtc2017,Giechaskiel2016}. DNSSEC relies on the distributed design of redundant DNS servers as well as caching. GNS stores records redundantly via replication in a Distributed Hash Table (DHT) and also uses a response caching mechanism to ensure availability.

\paragraph{Privacy and Confidentiality} Records are usually not considered confidential in a name system as its primary use-case is resource discovery. 
As such, namecoin and DNSSEC do not protect the contents of resource records or namespaces in any way. 
The blockchain-based design of namecoin in particular makes this property hard to satisfy, as all information is redundantly stored by all participants while DNSSEC suffers from a design weakness that results in a privacy issue regarding namespace enumeration\footnote{https://dnscurve.org/espionage2.html, accessed 12/27/2016} that was fixed only recently~\cite{goldberg2015nsec5}.
GNS, however, has a feature called \textit{query privacy} that protects against namespace enumeration and also ensures the confidentiality of records under certain circumstances.

\section{Implementation}
\label{sec:design}
In the following, we discuss the details of our ABD prototype implementation on top of GNS.
It is a DHT-based name system built on the GNUnet peer-to-peer framework\footnote{https://gnunet.org/gns, accessed 2/9/2017}. The underlying DHT provides reasonable performance, censorship resistance as well as some anonymity properties~\cite{evans2011r5n}. 
The design of GNS is inspired by SDSI and allows namespace owners to delegate authority over names in local namespaces to other participants. 
We have found GNS to be the best match for an ABD system, as it has the strongest attacker model of all name systems that the authors are aware of.
Namespaces in GNS are uniquely identified by a public-private key pair $(P,x)$
and referred to as \emph{identities} of the owner.


Creating a delegation to another namespace is effectively creating a direct trust-relationship.
Trust in GNS is established out-of-band through a key exchange. Records in
a namespace that is not trusted directly can be resolved if there is an
indirect, delegated trust path to that namespace.

\begin{figure*}[h!]
	\centering
	\includegraphics[width=0.7\linewidth]{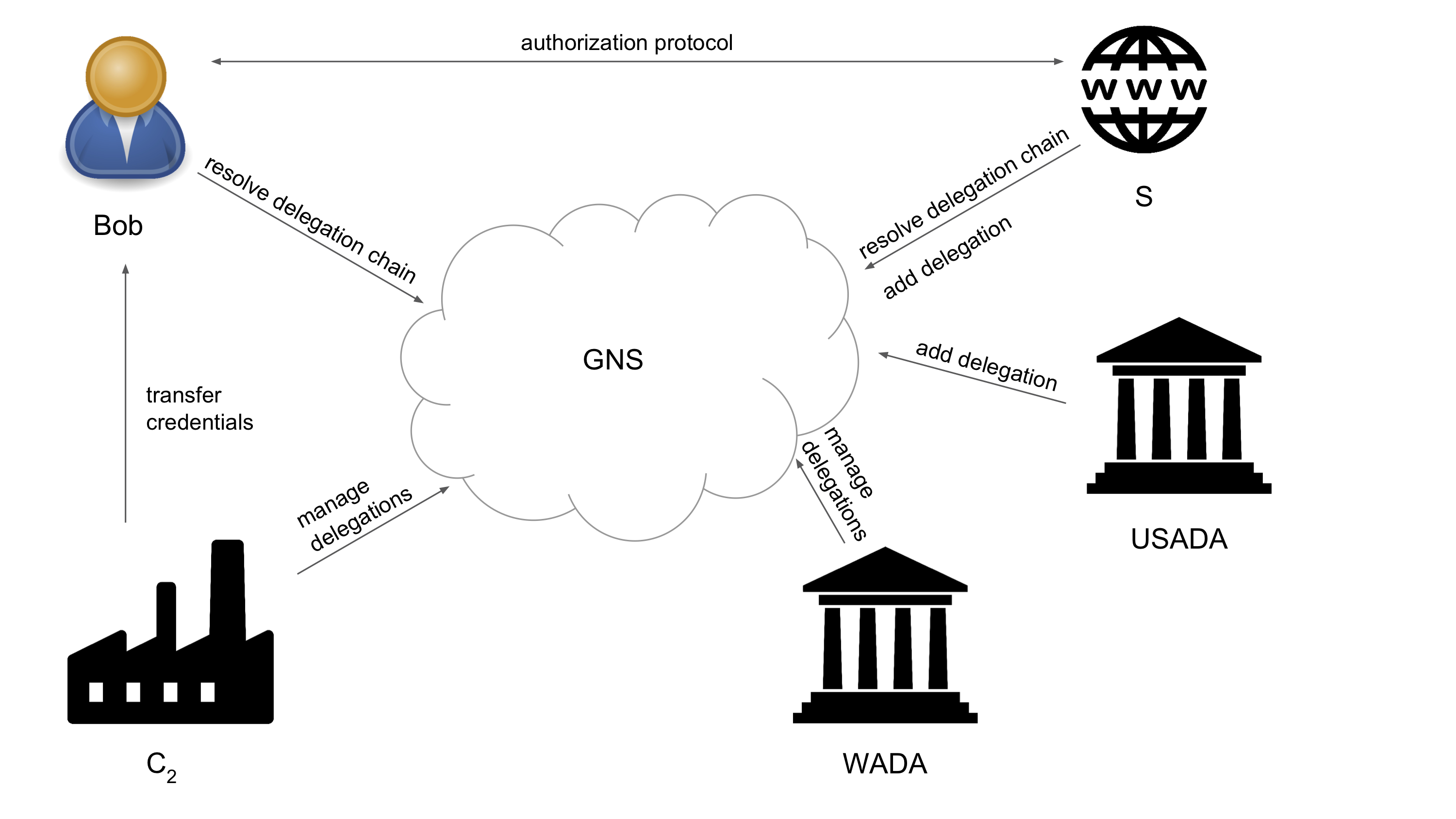}
	\caption{Scenario overview.}
	\label{fig:scenario_overview}
\end{figure*}

\subsection{Architecture}
To realize our proposed design on top of GNS, we have implemented a GNUnet service that is divided into three components: The \textit{Delegation Resolver}, the \textit{Delegation Manager} and the \textit{Credential Manager}. 
Each component reflects one or more functional parts of our design. All participants in the ABD system run a GNUnet peer, including our ABD service components.
While all participants require the Delegation Resolver functionality, only verifiers require the Delegation Manager to manage attribute delegations, and only credential subjects and issuers require the Credential Manager for the issuance and storage of attribute credentials.
The functionality of all components is exposed in REST APIs. 

\paragraph{\textit{Delegation Resolver}} This component includes the functionality for credential collection and chain verification as discussed in Section~\ref{sec:abd_auth}. The delegation chain discovery algorithm uses GNS to resolve delegations.
\paragraph{\textit{Delegation Manager}} The addition and removal of attribute delegations between GNS identities is implemented in the Delegation Manager component. 
In particular, it is used to manage ``ATTR'' delegation records that are persisted in the local namespace of the delegation issuer. The GNS service component periodically and automatically publishes those in the DHT. Only then are the delegations resolvable by other participants.
\paragraph{\textit{Credential Manager}} While our design allows us to use any kind of attribute-based credential, the Credential Manager enables users to issue simple credentials in our prototype implementation. A subject manages his credentials locally in a namespace using credential records with type ``CRED''. 
Credentials records contain the credential issuer public key, the subject public key, the expiration date, the attribute that is asserted as well as a signature. 
Credentials created using the Credential Manager are transferred to the subject in a JSON format. The JSON credential is converted by the subject's Credential Manager to a credential record and stored in a local namespace. 

\subsection{Authorization Protocol}
The authorization protocol introduced in Section~\ref{sec:abd_auth} is implemented on top of the system proposed by \cite{schanzen2016dpm}.
It allows a relying party (RP) to securely request attributes from a user and verify his identity at the same time, which is similar to an OpenID-Connect authorization flow~\cite{website:oidc} but without a trusted identity provider.
However, in the original design, the attributes that are provided by the user are not asserted by any third party.

We use the protocol to allow an RP to request delegated attributes that are required in a policy $P$.
For this, we extend the implementation to allow the user to present sets of credentials in an authorization response.
The response allows the RP to verify that the user is in possession of credentials that assert him a certain attribute through delegation.
Our modifications to the protocol are minor in that we simply redefine the interpretation of a requested attribute to be a delegated attribute by the RP.
The response is a JSON Web Token (JWT) that contains credential sets instead of self-signed attributes. As the technical basis of the protocol remains untouched, the security assurances and proofs presented in \cite{schanzen2016dpm} still hold.

Figure~\ref{fig:scenario_overview} provides an overview of the implemented scenario.
To integrate the scenario into our implementation, we set up five GNUnet peers. On each peer we created an identity that represents one of our entities including the service $S$, $WADA$, $USADA$, $C_2$ and $Bob$.
We set up all delegations as defined in Section~\ref{sec:abd} for each respective identity by using the Delegation Manager. 
We implemented a simple credential issuing website that allows $Bob$ to retrieve credentials issued by $C_2$ using the Credential Manager. Further, we created a demo website that represents the service $S$ that wants to provide restricted functionality to DCOs. This service initiates the authorization protocol and uses the Delegation Resolver for attribute verification.
\clearpage

\section{Conclusions}
\label{sec:Conclusion}
We argue that most of today's distributed applications unnecessarily rely on central trusted identity providers that require ultimate trust by all parties and amass piles of private information, including attribute values and activity profiles. 
In this paper, we show that such trusted third parties can be replaced by decentralized attribute-based delegation (ABD) mechanisms that have been proposed by other authors before at mostly theoretical levels.

Our contribution is the continuation of their work and the transfer into a real-world use case by means of designing and implementing a practical ABD system. 
 In particular, we point out that secure name systems provide a suitable basis for the implementation of ABD systems due to the inherent feature of authority delegation, and argue that systems like GNS, which are designed against strong attacker models and feature security properties like query privacy, are most suitable for realizing ABD systems.
We show how all necessary features of an ABD system can be implemented on top of the GNU Name System and how the properties of GNS are leveraged to achieve the security requirements of an ABD system.
The respective implementation\footnote{\url{https://gnunet.org/git/gnunet.git/tree/src/credential}} on top of GNUnet and a demo application\footnote{\url{https://github.com/schanzen/gnuidentity-example-rp/tree/credential}, \url{https://github.com/schanzen/gnunet-webui/tree/credentials}} can be found online. 
It is reasonable to assume that the performance of our ABD system is mainly influenced by the caching strategy of the underlying name system, which makes responsiveness of the name system an important criterion when designing an ABD system.
As the GNU Name System is built on top of a DHT, this is a valid concern and should be evaluated further.


In future work, we consider enhancing our system with distributed trust negotiation instead of our basic authorization protocol, such as the one proposed by Li et al~\cite{li2003rt}. Additionally, we are planning on integrating our ABD system into an authorization framework that uses the UMA protocol for standardized authorization management and, at the same time, takes advantage of decentralized storage and evaluation of ABD-based policies.

\section*{Acknowledgment}
This work has been partially funded in the project PARADISE by the German Federal Ministry of Education and Research under the reference 16KIS0422.


%
\bibliographystyle{IEEEtran}
\bibliography{references}  

\end{document}